# High-temperature superconductivity in iron-based materials

## Johnpierre Paglione and Richard L. Greene

*Center for Nanophysics and Advanced Materials, Department of Physics, University of Maryland, College Park, MD 20742*

The surprising discovery of superconductivity in layered iron-based materials, with transition temperatures climbing as high as 55 K, has lead to thousands of publications on this subject over the past two years. While there is general consensus on the unconventional nature of the Cooper pairing state of these systems, several central questions remain - including the role of magnetism, the nature of chemical and structural tuning, and the resultant pairing symmetry - and the search for universal properties and principles continues. Here we review the progress of research on iron-based superconducting materials, highlighting the major experimental benchmarks that have been so far reached and the important questions that remain to be conclusively answered.

In February 2008, Hideo Hosono and co-workers reported the discovery of 26 K superconductivity in fluorine-doped LaFeAsO [1], marking the beginning of worldwide efforts to investigate this new family of superconductors. While rumors of a 50+ kelvin superconductor swirled around the 2008 APS March meeting in New Orleans, researchers in Japan and China were busy with experiments that would largely advance this field to its current status by raising $T_c$ values of LaFeAs(O,F) to 43 K by application of pressure [2] and then as high as 55 K by replacement of La by other rare earth elements [3]. Historically, the typically antagonistic relationship between superconductivity and magnetism has led researchers to avoid using magnetic elements - ferromagnetic in particular - as potential building blocks of new superconducting materials. Because elemental iron is strongly magnetic, the discovery of Fe-based superconductors (FeSCs) with high-$T_c$ was completely unexpected. This has opened a new avenue of research driven by the fact that our fundamental understanding of the origins of superconductivity needs significant improvement.

This Progress Article provides a summary update on some of the recent experimental results and an overview of the status of the field. Our goal is to highlight important experimental observations and theoretical perspectives that may lead to a consensus on the understanding of superconductivity in the FeSCs, leaving detailed overviews to existing reports [4]. The basic behavior of several classes of FeSCs is now known to be quite similar, so this review will cover universal properties but focus on the class of intermetallic FeSCs with the ThCr$_2$Si$_2$ (122) structure, pointing out any significant differences from the other systems.

## Crystal structures and tuning parameters

To date, five different structural classes of iron-based SCs have been found. These structures, shown in **Box 1**, all share a common layered structure based upon a planar layer of iron atoms joined by tetrahedrally coordinated pnictogen (P, As) or chalcogen (S, Se, Te) anions arranged in a stacked sequence separated by alkali, alkaline earth or rare earth and oxygen/fluorine "blocking layers." It is now widely thought that the interaction that leads to the high-$T_c$ superconductivity originates within these common iron layers, similar in nature to the common copper-oxide building block found in the copper-oxide (cuprate) high-$T_c$ superconductors [5]. As in the cuprates, chemical substitution also plays a key role in inducing the superconducting phase in iron-pnictides. However three key differences are found, (1) in the arrangement of pnictogen/chalcogen anions above and below the planar iron layer (see **Box 1**) as opposed to the planar copper-oxygen structure of the cuprates; (2) in the ability to substitute or dope directly into the active pairing layer; and (3) in the metallic (rather than insulating) nature of the parent compounds. It is these traits, together with the similar interplay of magnetism and superconductivity, that mark the iron-pnictides and cuprates as distinct, but closely related, superconducting families. The phase diagram of the iron-based SCs is in fact strikingly similar to several other classes of unconventional SCs, including the cuprates, organics and heavy-fermion SCs, all believed to harbor unconventional (non-phonon-mediated) pairing mechanisms (for a perspective, see [6]). While the mediator of pairing in these systems remains officially unidentified, a large amount of circumstantial evidence points to magnetic spin fluctuations: in all cases magnetism must be suppressed, either by pressure or doping, before optimal bulk phase superconductivity appears. Although their more metallic nature may place them closer to the heavy-fermion systems (metallic magnetism) than to cuprates (Mott insulator physics), the striking resemblance of interpenetrating ground states in these systems deserves strong attention in devising a generalized theory of unconventional superconductivity.

The generic phase diagram of the FeSC systems can be produced by manipulating the chemical or structural properties, using either chemical doping/substitution or applied external pressure to drive an antiferromagnetic (AFM), non-superconducting parent compound to a superconducting (SC), non-AFM state. A compilation of experimental phase diagrams is presented in **Fig. 1** for the Ba-based "122" system, to date the most widely studied of the five families and widely thought to capture the main traits of all FeSCs. In BaFe$_2$As$_2$, the systematic substitution of either the alkaline earth (Ba), transition metal (Fe) or pnictogen (As) atom with a different element almost universally produces the phase diagram presented in **Fig. 1a**, composed of an AFM state that is suppressed with substitution and a SC phase that is more or less centered near the critical concentration where AFM order is destroyed. This is somewhat different from the known behavior of F-doped "1111" systems such as LaFeAsO$_{1-x}$F$_x$ [7], where AFM and SC phases are completely separated as a function of doping and do not overlap. However, the coexistence of AFM and SC phases such as reported for SmFeAsO$_{1-x}$F$_x$ [8] is believed to likely be the more intrinsic property of the generic FeSC phase diagram, motivating efforts to study the 122 systems in great detail. The quantitative similarity between phase diagrams produced by substitutions involving both obvious (*i.e.*, K$^{1+}$ for Ba$^{2+}$) [9] and subtle (*i.e.*, Co-3$d^7$ for Fe-3$d^6$) [10] charge doping, as well as nominally isovalent (P-3$p^3$ for As-4$p^3$) [11] substitutions, is enticing due to the implied versatility of chemical tuning parameters available to experimentalists for studying these systems. Furthermore, it promotes the idea that simple charge doping, as understood to be the case in the cuprates, is not the sole factor in determining the phase boundaries of these systems and that structural tuning may play a role.

However, subtleties in the electronic structure of these materials, as discussed below, make the situation more complex than that of a simple structural tuning effect. This is highlighted by the sensitivity of the superconducting phase to particular choice of ion substituent. For example, superconductivity in the 122 materials, first shown to occur by Co substitution for Fe in SrFe$_2$As$_2$ [12], can be stabilized by several types of *d*-metal substitution. This includes the use of any elements in the Fe, Co and Ni columns (except, to date, Os) [13], but excludes Cr [14], Mn [15] and Cu [16] which all act to suppress magnetism without stabilizing a SC phase. It is thought that these latter anomalous cases arise for varying reasons [14,16-18] to do with the disfavorable manipulation of Fe bonding and magnetism, giving clues regarding the correct distinction between charge doping and chemical substitution.

Pressure tuning is less well understood. In some cases this powerful control parameter is aligned with its chemical substitution counterpart. For instance, in Ba$_{1-x}$K$_x$Fe$_2$As$_2$ a good overlap exists between lattice parameter variation by applied pressure or K substitution [19], allowing conclusions about the roles of lattice structure versus charge doping to be made. In contrast, in pressure experiments on BaFe$_2$As$_2$, differing experimental conditions impose variations from true hydrostatic conditions, making it difficult to generically compare phase diagrams obtained via applied pressure versus chemical substitution. **Fig. 1b** presents a comparison of five studies [20-24] on our model 122 system using differing techniques, showing that AFM order is suppressed in a manner similar to chemical substitution in all cases shown but with differing rates. Moreover, the pressure range where the superconducting dome is located also varies for each experiment. This is likely due to the fact that the compressibility of BaFe$_2$As$_2$ is highly anisotropic [25], imposing a sensitivity to non-hydrostatic pressure conditions that may alter the evolution of the phase diagram under differing experimental conditions. Such a scenario was recently shown conclusively in a comparison of pressure experiments using the same crystals but different levels of hydrostaticity [26].

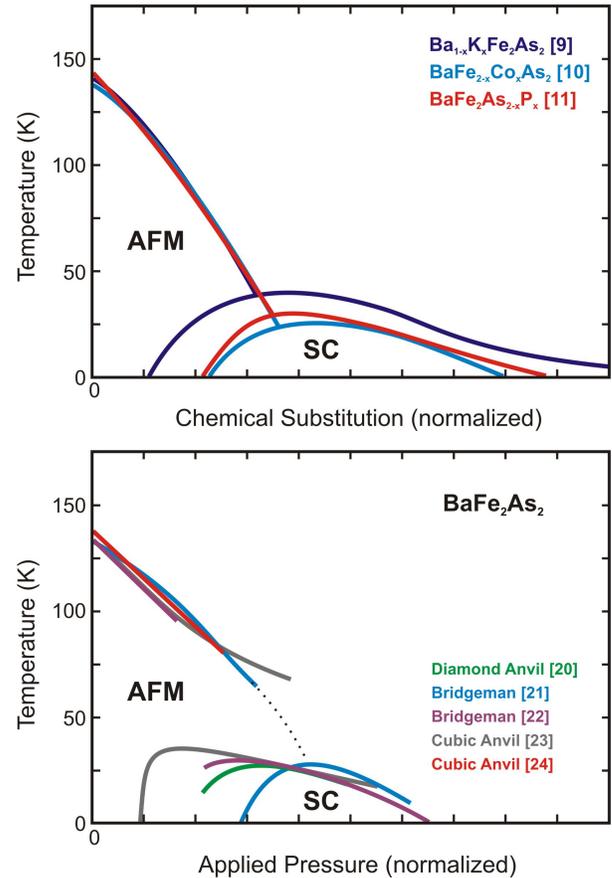

**Figure 1**: **Experimental phase diagrams of the BaFe$_2$As$_2$ system. (top)** Chemical substitution phase diagram of BaFe$_2$As$_2$ system, shown for K [9], Co [10] and P [11] substitutions, with the amount of chemical substitution (x) normalized to overlap the descent of the antiferromagnetic transition for simplified comparison. **(bottom)** Pressure phase diagram for BaFe$_2$As$_2$ as a function of external pressure applied under various levels of hydrostaticity, using diamond anvil cell [20], Bridgeman [21,22], and cubic anvil cell [23,24] techniques. Note that the pressure axis is normalized to overlap the descent of the antiferromagnetic transitions for each experiment for simplified comparison.

In one extreme case involving pressure-tuning of CaFe$_2$As$_2$, an instability to a structural phase transition (the so-called "collapsed tetragonal" phase) imposes a severe sensitivity to anisotropic strain conditions, with a pressure-induced SC phase only present when non-hydrostatic conditions are imposed [27]. While it remains unclear what role structure plays in stabilizing superconductivity in CaFe$_2$As$_2$, some theoretical ideas [28] suggest As-As bonding to be the key ingredient. The sensitivity to hydrostaticity certainly implies that a strain mechanism is at work, possibly similar to what causes the intermittent appearance of 20 K superconductivity in undoped, unpressurized 122 parent compounds [29]. Indeed, strain effects have been identified in studies of twin domain boundaries of $BaFe_{2-x}Co_xAs_2$ using scanning SQUID microscopy, where an enhanced susceptibility at twin boundaries has been associated with an enhanced superfluid density [30]. More generally, one of the distinguishing features of the FeSC materials is the fact that the generic phase diagram can be experimentally tuned by any of several different means that allow for precise control of structural parameters, disorder location, chemical bonding and density. This is one of the key properties that has allowed for a rapid but in-depth understanding of these materials. In due time, controlled experimental comparisons - for instance of Hall effect (carrier density) under pressure versus doping, comparison of different chemical substitution series, and further understanding of the local nature of chemical substitution - will help pinpoint the important tuning parameters for these systems.

**Electronic structure**

With an interplay of magnetic and electronic interactions likely playing an integral role in determining the shape of the phase diagram of all FeSC systems, much work has gone into elucidating the magnetic and electronic structures of these materials. In general, these materials are well described as consisting of two-dimensional (2D) metallic sheets derived from Fe $d$- states hybridized with As $p$-orbital derived bands, sitting in a quasi-ionic framework composed of rare earth, oxygen, alkali or alkaline earth "blocking layers" (see **Box 1**). This arrangement unites to produce a metallic material with nominal Fe valence of 2+, low carrier concentration and high electronic density of states dominated by Fe $d$-states [31]. Based on this FeAs-layered framework, the electronic band structure has been calculated using the local density approximation (see [32] for an overview), showing that the electronic properties are dominated by five Fe $d$-states at the Fermi energy, with a Fermi surface (FS) consisting of at least four quasi-2D electron and hole cylinders (see **Box 2**). These consist of two hole pockets centered at the Brillouin zone (BZ) center and two electron pockets centered at $(0, \pm\pi)$ and $(\pm\pi,0)$. Two nonequivalent As positions (staggered above and below the Fe lattice) result in folding of the BZ to include two Fe atoms per unit cell and to put the electron pockets at $(\pm\pi,\pm\pi)$, the same direction of AFM ordering vector as discussed below. A fifth hole band is also proposed to sit at $(0,\pm\pi)$ in the folded BZ, and its presence may be very sensitive to structural details [33].

The qualitative agreement with experiment is remarkably good, as shown by several angle-resolved photoemission spectroscopy (ARPES) and quantum oscillations measurements. Initial ARPES studies on LaFePO [34], $NdFeAsO_{0.9}F_{0.1}$ [35], and $Ba_{0.6}K_{0.4}Fe_2As_2$ [36] all confirmed the predicted band structure composing of hole pockets at the BZ center and electron pockets at the BZ corners. Quantum oscillations experiments performed on $CaFe_2As_2$ [37], $SrFe_2As_2$ [38] and $BaFe_2As_2$ [39] parent compounds, as well as on P-doped $BaFe_2As_2$ [40], are also consistent with this basic structure. As usual, the devil is in the details: simple 2D models may not capture subtle but important details of the true band structure, and three-dimensional (3D) aspects may have important implications for magnetism and superconductivity. While ARPES is a surface probe, it is capable of probing $\mathbf{k}_z$ dispersion to a limited extent, and several studies [41-43] have reported strong modulation along the $\mathbf{k}_z$ direction at the BZ center in agreement with theory [44].

While being another area of active debate, the effect of hole/electron doping on the electronic structure is fairly well captured by a rigid band picture: the basic FS topology is kept with both electron (i.e. in $BaFe_{2-x}Co_xAs_2$ [45]) and hole (i.e. in $Ba_{1-x}K_xFe_2As_2$ [46]) doping, with the size of FS pockets changing accordingly and with reasonable continuity observed when crossing between each case [47]. Once again, however, it appears that charge doping is not the only mechanism by which this tuning can be achieved. For instance, isovalent substitutions would not be expected to mimic charge-doping effects, however the nominally isovalent substitutions of P for As [48] and Ru for Fe [49] in $BaFe_2As_2$ indeed cause substantial changes to the Hall coefficient, even changing sign in the latter case. Moreover, a comparable scaling of FS pocket size has been shown to occur as a function of isovalent As/P substitution [40], and density functional theory (DFT) calculations suggest that the main effect of charge doping is not on the density of states, but rather on the disruption of nesting and the relative size of electron and hole pockets [44]. Beyond isovalent substitution, a DFT study of the local nature of $d$-electron "dopant" ions (including the likes of Co and Ni) concluded that the extra electrons actually remain predominantly localized at the substituent site, and may act mainly to disrupt nesting properties rather than changing charge density [50].

More profound changes in band structure may also be at play. ARPES studies of $BaFe_{2-x}Co_xAs_2$ provide indications of a possible Lifshitz transition where hole pockets disappear, allowing superconductivity to thrive [51]. A change in effective dimensionality across the AFM transition - with development of a 3D ellipsoid in $CaFe_2As_2$ below $T_0$ - has been suggested to be important for SC [42], but is in conflict with the observation of significant 3D structure in $BaFe_{2-x}Co_xAs_2$ outside of the AFM phase [43]. Furthermore, a qualitative change in band structure appears to occur as a function of Co-doping at the onset of the SC phase, as observed in ARPES [51] and Hall data [52], and is also supported by analysis of transport data [53,54].

The connection between structural details of FeSC materials and their seemingly sensitive electronics is an important aspect of superconductivity in the FeSCs. An empirical relation between the tetrahedral bond angle of the As-Fe-As layer and $T_c$ values for different FeSCs was recognized early on, with optimal $T_c$ values suggested to be

## Box 1: The iron-based superconductor family

Iron, one of the most common metals on earth, has been known as a useful element since the aptly named "iron age". But it was not until recently that, when combined with elements from the peculiar group 15 and 16 columns of the periodic table (named, respectively, the "pnictogens" after the greek verb for choking, and "chalcogens" meaning "ore formers"), iron-based metals have been shown to readily harbor a new form of high-temperature superconductivity. This general family of materials has quickly grown to large size, with well over 50 different compounds identified to show a superconducting transition that occurs at temperatures that approach 60 K, and includes a plethora of different variations of iron- and nickel-based systems. To date, five unique crystallographic structures have been shown to support superconductivity. As shown in **Fig. B1a)**, these structures all possess tetragonal symmetry at room temperature and range from the simplest α-PbO-type binary element structure, to more complicated quinternary structures composed of elements that span the entire periodic table.

The key ingredient is a quasi-two-dimensional layer consisting of a square lattice of iron atoms with tetrahedrally coordinated bonds to either phosphorus, arsenic, selenium or tellurium anions that are staggered above and below the iron lattice to form a checkerboard pattern that doubles the unit cell size, as shown in **Fig. B1b)**. These slabs are either simply stacked together, as in FeSe, or are separated by spacer layers using alkali (*e.g.*, Li), alkaline earth (*e.g.*, Ba), rare earth oxide/fluoride (*e.g.*, LaO or SrF) or more complicated perovskite-type combinations (*e.g.*, $Sr_3Sc_2O_5$). These so-called "blocking layers" provide a quasi-two-dimensional character to the crystal since they form atomic bonds of more ionic character with the FeAs layer, while the FeAs-type layer itself is held together by a combination of covalent (*i.e.*, Fe-As) and metallic (*i.e.*, Fe-Fe) bonding.

In the iron-based materials, the common FeAs building block is considered a critical component to stabilizing superconductivity. Because of the combination of strong bonding between Fe-Fe and Fe-As sites (and even interlayer As-As in the 122-type systems), the geometry of the $FeAs_4$ tetrahedra plays a crucial role in determining the electronic and magnetic properties of these systems. For instance, the two As-Fe-As tetrahedral bond angles appear to play an crucial role in optimizing the superconducting transition temperature (see main text), with the highest $T_c$ values found only when this geometry is closest to the ideal value of ~109.47°.

Long range magnetic order also shares a similar pattern in all of the FeAs-based superconducting systems. As shown in the projection of the square lattice displayed in **Fig. B1b)**, the iron sublattice undergoes magnetic ordering with an arrangement consisting of ferromagnetically arranged spins along one chain of nearest neighbours within the iron lattice plane, and antiferromagnetically arranged along the other direction. This is displayed on a tetragonal lattice in the figure, but actually only occurs after these systems undergo an orthorhombic distortion as explained in the main text. In the orthorhombic state, the distance between iron atoms with ferromagnetically aligned nearest neighbor spins (highlighted in Fig. B1b) shortens by approximately 1% as compared to the perpendicular direction.

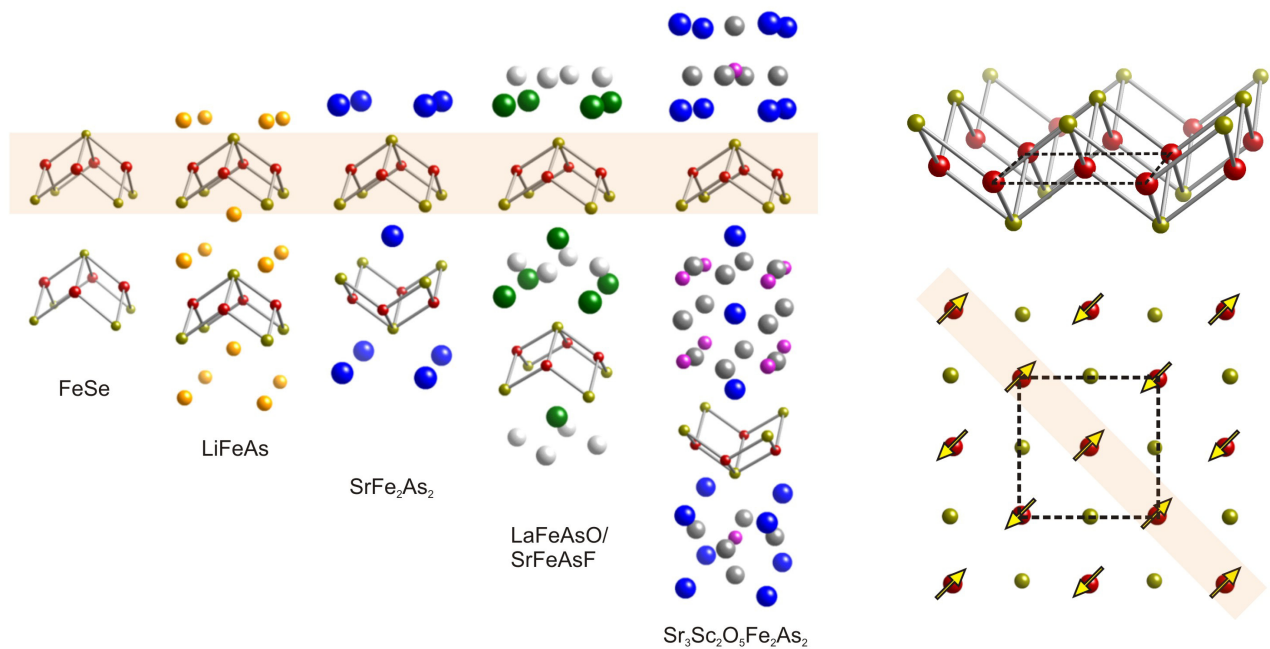

**Figure B1: Crystallographic and magnetic structures of the iron-based superconductors. a)** The five tetragonal structures known to support superconductivity. **b)** Depiction of the active planar iron layer common to all superconducting compounds, with iron ions shown in red and pnictogen/chalcogen anions shown in gold. The dashed line indicates the size of the unit cell of the FeAs-type slab that includes two iron atoms due to the staggered anion positions.

dependent upon ideal tetrahedral geometry [55]. This has important implications, both theoretically and in practical terms: a close relationship between structure and superconductivity, direct or indirect, places constraints on both the theoretical understanding of the pairing interaction and the promise of superconductors with higher $T_c$ values. While it appears this relationship is not universal (a notable exception includes $CsFe_2As_2$, with $T_c$=2.6 K and As-Fe-As bond angle very close to the ideal value of 109.47° [56]), it remains true that the highest-$T_c$ FeSCs are still clustered close to the ideal tetrahedron geometry. A proposed sensitivity of electronic structure and/or magnetic interactions to details of the internal structure of the Fe-As layers is likely relevant to unraveling this puzzle. For example, calculations of the dependence of nesting in the band structure, and consequently pairing in a spin fluctuation-mediated scenario, show a sensitivity of an extra (third) hole pocket to pnictogen height above the Fe plane with consequences for pairing symmetry [33]. Recent experimental studies of the relationship between $T_c$ and anion height in the related Fe(Se,Te) binary superconductors, as a function of pressure [57] and epitaxial strain of thin films [58] have allowed for more systematic studies of this intriguing relationship.

**Magnetism**

The nature of magnetism in the FeSC parent compounds is a hotly debated topic, largely due to its implications for the pairing mechanism: the electronic structure suggests that the same magnetic interactions that drive the AFM ordering also produce the pairing interaction for superconductivity [59]. As predicted before experiments [60], AFM order in all FeAs-based superconducting systems is found to have a wavevector directed along $(\pi,\pi)$ with a real-space spin arrangement consisting of AFM stripes along one direction of the Fe sublattice and FM stripes along the other (see **Box 1**), with an ordered moment typically smaller than one Bohr magneton [61,62]. Early on, theoretical calculations found these materials to lie near a Stoner instability, suggesting inherent itinerant magnetism [31], which could potentially explain the consistent overestimates of the ordered moment size using first-principles calculations. However, this ordering arrangement also falls naturally out of a local-moment "$J_1$-$J_2$" model with nearest ($J_1$) and next-nearest ($J_2$) neighbor exchange interactions such that $J_1<2J_2$ [63], and hence a lot of effort has gone into determining whether such a localized model holds any validity for a system that is clearly metallic.

On the itinerant side, most models focus on a spin-density wave (SDW) instability of the Fermi surface. While there are not many direct observations of a SDW energy gap, optical studies of $BaFe_2As_2$, $SrFe_2As_2$ and $EuFe_2As_2$ have indeed shown evidence for gapping of the Fermi surface below $T_N$ [64]. It is widely thought that the SDW instability arises from nesting of two FS pockets by a large $\mathbf{Q} = (\pi,\pi)$ vector that is commensurate with the structure. This vector corresponds to the magnetic ordering vector measured throughout the FeAs-based parent compounds [61,62] as well as that for magnetic fluctuations in the superconducting compounds [65]. There is varied, but good, evidence for $(\pi,\pi)$ FS nesting across the FeSC family, as told by ARPES measurements of $BaFe_{2-x}Co_xAs_2$ [47,66] and $Ba_{1-x}K_xFe_2As_2$ [36,46], and quantum oscillations measurements in LaFePO [67] and overdoped $BaFe_2As_{2-x}P_x$ [68]. In addition, the closely related pnictide material FeTe also exhibits nesting in the same $(\pi,\pi)$ direction, even though its magnetic ordering vector is shifted by 45° [69].

In the parent compounds, there are differing results regarding the change in band structure through the magnetic transition as interpreted via ARPES data, but bulk measurements of electronic structure are more consistent. All studies [37-39] observe small FS pockets consistent with band structure calculations involving a conventional band-folding picture of the AFM ground state involving quasiparticles in a conventional SDW model. While the general picture is well fit with a SDW model, there are problems with matching to a purely itinerant scenario. In particular, the increased conductivity found in SDW state is not expected if a portion of the carriers become gapped (although a tradeoff between carrier number and scattering can explain the reversal observed in doped materials where the conductivity decreases below $T_N$). Also, there is significant, non-trivial reconstruction of the FS below $T_N$ that goes beyond simple band-folding (*e.g.*, see ARPES data on $EuFe_2As_2$ [70]). Theoretically, these concerns can be taken into account in an itinerant scenario by considering details of the multi-band structure [71,72], with the non-trivial band folding and metallic nature of the FeSCs likely a result of partial gapping of the FS.

"Locally", neutron scattering measurements of $CaFe_2As_2$ have revealed magnetic spin wave excitations to live up to ~200 meV, with low energy modes that can be phenomenologically fit to a local-moment 3D Heisenberg model [73,74]. One study [73] observes strong damping at high energies consistent with the presence of particle-hole excitations, while another [74] does not observe this damping but argues that a large in-plane anisotropy of local-moment interactions, together with incompatibility of a spin-1/2 model with even-numbered electrons in Fe-$3d^6$, necessitates the presence of an itinerant component. Above all, it is safe to say that these materials are "itinerant by definition" [61], since all Fe-based bands cross the Fermi energy and thus intimately involve conduction electrons in forming the magnetic state. But one should not exclude the applicability of less simplistic forms of magnetic order, including orbital order [75], coexistent [76] itinerant electrons and local moments, or an admixture of both local coupling and itinerant one-electron interactions [77].

Regardless of the exact nature of magnetic order, it is clear that magnetostructural coupling is prevalent throughout the FeSC family. A structural phase transition ($T_0$) - from tetragonal at high temperatures to orthorhombic at low temperatures - is almost always coupled to the magnetic transition at $T_N$ (making life more difficult for establishing ideal conditions in a pressure experiment, for instance due to complications arising from twinning domains). Theoretically, this magnetostructural coupling is understood to be a consequence of the magnetic interactions [63,32], implying the existence of the structural transition is merely circumstantial. A peculiarity of the coupled transitions is that, aside from the case of the 122 parent compounds where $T_N$ and $T_0$ coincide, the structural and magnetic phase transitions are positioned at *different* temperatures in 1111 compounds, with the structural transition actually preceding the magnetic transition [78]. A comparison of this phenomenon in samples of CeFeAsO with differing levels of disorder (polycrystal, cluster, single crystal) has shown the splitting to decrease (from 18 K to 6 K) with improved sample quality [79]. Considering that the splitting between $T_N$ and $T_0$ is established in 122 compounds only after chemical substitution [80], this would lead to the conclusion that disorder is the cause of the splitting and that magnetostructural coupling in the cleanest systems should result in coincident transition temperatures. However an intrinsic mechanism for this splitting cannot be ruled out, especially considering the early prediction of an Ising phase transition that should precede $T_N$ [81]. It will thus be interesting to see if experiments that tune effective dimensionality and/or magnetic coupling will reveal further insight into the nature of the splitting. Finally, a direct coupling between structural distortion and magnetism has also been shown in P-doped CeFeAsO, where neutron scattering measurements reveal an ordered AFM moment of Fe to be linearly proportional to orthorhombicity (i.e. lattice constant ratio *a/b*) as a function of P concentration [82]. While seemingly in line with strong magnetostructural coupling discussed above, this result remains highly enigmatic owing to the apparent contradiction with first-order Landau theory, which forbids a linear relationship between these vector and scalar order parameters. Because such a relation was also first shown via x-ray and muon spin resonance in $SrFe_2As_2$ [83], this strange but apparently universal coupling remains as an important topic in our understanding of magnetostructural coupling in the FeAs-based materials.

**Superconducting state—pairing symmetry**

As for the high-$T_c$ cuprates, the fundamental mechanism which causes the high temperature superconductivity in the FeSCs is a question of primary importance. In both cases the experimental evidence to date favors an unconventional pairing mechanism closely tied to magnetism. Although the exact nature of the pairing is not known in either system at present, many experiments aimed to determine the pairing symmetry have been performed. For the cuprates, the experimental evidence favors a singlet *d*-wave symmetry that involves a change in sign of the superconducting order parameter (OP) phase at nodal points situated at the Fermi energy ($E_F$) and directed along $(\pi,\pi)$ in the simple 2D cuprate band structure. For the FeSCs, the initial measurements [36,84] probing the OP symmetry pointed to a *fully gapped* OP consistent with a fully symmetric *s*-wave symmetry. In comparison to cuprates and other magnetically mediated superconductors, this came as a surprise. However, the OP symmetry of FeSCs was in fact predicted theoretically to have *s*-wave symmetry, but with a sign change that occurs between different bands in the complex multi-band electronic structure. This is the so-called "$s\pm$" symmetry, calculated before experiments [60] (see [59] for details), that will be predominantly discussed below in the context of experimental work.

Experiments that probe the symmetry of the SC phase provide important information about the energy and momentum dependence of Cooper pairing, and are therefore pivotal to helping elucidate the pairing mechanism in this new class of high-Tc superconductors. Due to the vastness of the iron-pnictide family and the nature of chemical substitution, one limitation to date is that many experiments have been performed on different systems or different chemical compositions of the same crystalline system, and thus make it difficult to compare. However there is surprisingly good consistency, allowing general conclusions to be drawn from several experiments. For instance, NMR experiments were quick to determine from Knight shift measurements that the SC state spin symmetry is likely singlet [85,86], implying an even OP symmetry (i.e. *s*-wave, *d*-wave, etc). To date, this experiment has been done on the main members of the FeSC family, including samples of the 1111 [85], 122 [86] and 11 [87] systems, so it is reasonable to assume that SC in the iron-pnictides is universally spin-singlet.

Determining the nature of the orbital OP symmetry, however, is much more complex and is currently the focus of most research. Theoretically, due to the multi-orbital nature of the FeSCs and their nesting properties, anisotropic (extended) *s*- and *d*-wave states are nearly degenerate [88], making it difficult to identify the underlying symmetry even if experiments determine the presence of nodes in the gap symmetry. Experiments to probe the orbital symmetry usually probe either the amplitude or the phase of the order parameter. For the cuprates, the phase experiments gave the most convincing evidence for a sign-changing *d*-wave ($d_{x2-y2}$) symmetry [89]: corner junction tunneling and SQUID experiments, observation of half-integer flux quantum from polycrystalline materials or bicrystal *ab*-plane oriented films, and absence of *c*-axis Josephson effect with an *s*-wave SC, were all taken as evidence for *d*-wave symmetry.

For the FeSCs, only a few phase experiments have been done to date. In polycrystalline 1111 no half-integer flux was found by a scanning SQUID experiment [90] and a robust *c*-axis Josephson effect was found between Pb and K-doped $BaFe_2As_2$ [91], which rules out a predominant *d*-wave symmetry in these materials at the measured K concentrations. However, to distinguish conventional *s*-wave symmetry from $s\pm$ is not as straightforward as for *d*-wave. Several proposals for tunneling/SQUID experiments have been made [92], but all require junction geometries more difficult to prepare than a corner junction, and no experimental results have so far been reported. One new type of phase experiment has been reported [93], where half-integer flux quantum jumps were observed in a loop formed by Nb and polycrystalline $NdFeAsO_{0.88}F_{0.12}$. These jumps were interpreted as arising from π phase shifts at a few polycrystal boundaries as current is passed through the sample, which could only occur if the symmetry is a sign-changing $s\pm$. A $d$-wave symmetry was ruled out because many more jumps would be expected in this case, and these were not observed by the scanning SQUID experiment [90]. Another phase sensitive experiment, just reported [94] on a single crystal of Fe(Se,Te), uses spectroscopic-imaging scanning tunneling microscopy to determine the sign of the OP gap by the magnetic field dependence of the quasi-particle scattering strength. Because of the coherence factors, the scattering amplitudes are different for sign-changing and sign-constant symmetries, as found in similar experiments on the cuprates [95]. In Fe(Se,Te), Hanaguri *et al.* found that the sign of the gap is reversed between the electron and hole pockets. This, along with an absence of low-energy quasiparticle excitations in the tunneling gap (as would be seen from the nodes in a $d$-wave SC), strongly suggests an $s\pm$ symmetry. Together, all of these results favor the $s\pm$ symmetry in the FeSCs, but definitive phase experiments on more materials are needed to conclusively settle the case.

Neutron scattering has been instrumental in helping clarify not only the magnetic properties [61], but also the interplay of magnetism and superconductivity in the FeSCs. For instance, strong evidence now exists for the competitive nature of AFM and SC coexistent phases in $BaFe_{2-x}Co_xAs_2$, as determined by the observation of a reduction of the static Fe moment upon entering the SC state [96]. A systematic study of the same nature [97], comparing experimental results to a mean field Ginsburg-Landau model, found this behavior to be inconsistent with a non-sign-changing $s$-wave symmetry, providing further support for $s\pm$ symmetry.

The observation of a collective magnetic resonance mode that appears below the SC transition temperature is a well known feature found by inelastic neutron scattering experiments in cuprate, heavy-fermion and, most recently, FeSC compounds. This feature, which appears to universally relate the energy of the resonance mode with the SC energy gap [98], has now been clearly observed in superconducting 122 [99] and 11 [100,101] compounds, providing important implications for both the pairing mechanism and the symmetry of the order parameter. As shown in **Fig.2**, the resonance observed in several experiments follows a very strong correlation with the SC state, as indicated by a scaling of the resonance energy with $T_c$. Because this relationship appears exclusively in nearly magnetic unconventional superconductors [98], it is considered either (optimistically) a strong signature of magnetically mediated pairing, or (pessimistically) an independent remnant of nearby magnetic order. In the first case, it is in fact possible to extract gap symmetry and phase information from the momentum dependence of the resonance by comparing to predictions for several differing gap symmetries [102], and therefore is a potent probe of the pairing mechanism. In the second case, the resonance appears as a consequence of the reduction of the rate of magnon decay via conduction electrons due to the opening of the SC gap at $T_c$.

While the true relation between the resonance and the pairing mechanism remains unclear, these opposing scenarios can be delineated by experimental determination of any fine structure of the resonance mode. A study by Bao *et al.* has addressed this issue in the 11 material $FeSe_{0.4}Te_{0.6}$ by studying the Zeeman splitting of the resonance in high magnetic fields [101]. They find that a triplet state emerges, providing indications of a two-particle bound state as opposed to a conventional magnon. This observation is very striking, especially because the resonance is associated with a ($\pi,\pi$) nesting vector even though long range order in FeTe has a *different* ordering vector at ($\pi,0$) [100]. However, the interpretation is complicated by the fact that the resonance is superimposed on incommensurate paramagnon excitations that disperse in energy as the resonance develops [103], making disentanglement of the splitting features in field difficult. Moreover, for experiments on BaFe$_{2-x}$Ni$_x$As$_2$, the absence of a splitting [104] as well as the anisotropy of the resonance [105] pose a challenge to the understanding of these results as "smoking gun" evidence of fluctuation-mediated pairing. Future measurements should be able to resolve this issue.

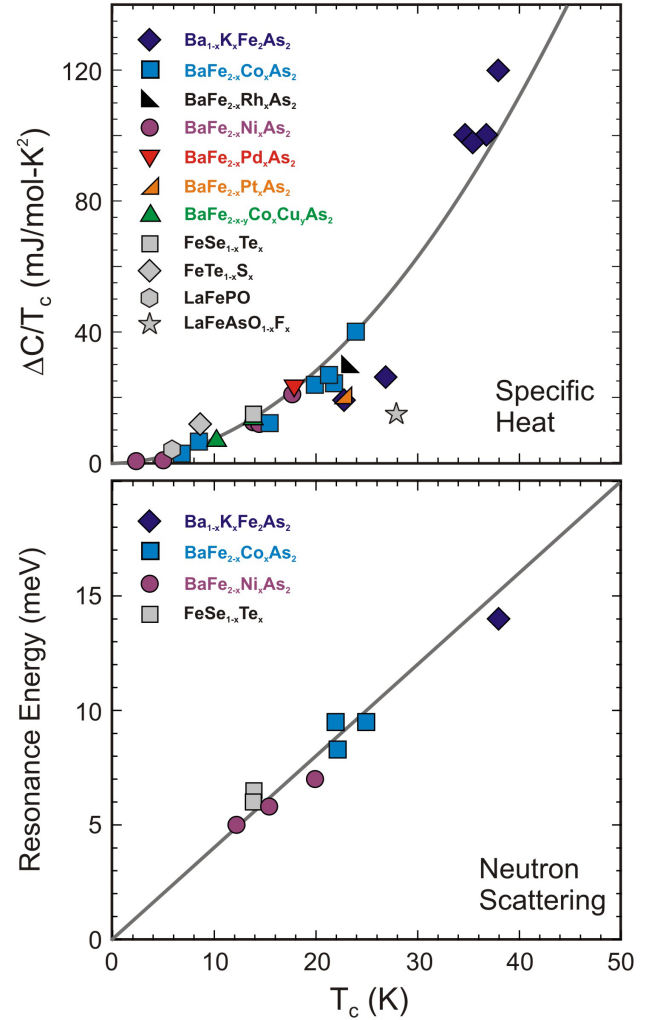

**Figure 2**: **Universal experimentally scalable quantities of FeAs-based superconducting materials.** (**top**) Absolute size of the measured jump in electronic specific heat at the superconducting transition in several members of 11, 122 and 1111 FeSC compounds plotted as a function of $T_c$. Originally identified in Co- and Ni-doped $BaFe_2As_2$ [136] and subsequently verified in K- [137], Co- [131,138], Rh-, Pd- [139] and Pt-doped [13] $BaFe_2As_2$ systems, as well as $FeTe_{0.85}S_{0.15}$ [140], $RFePO$ [141,142], $FeSe_{0.5}Te_{0.5}$ [143] and $LaFeAsO_{0.89}F_{0.11}$ [144] superconductors, this thermodynamic quantity appears to scale quadratically with $T_c$ in all cases, as demonstrated by the solid line fit to a quadratic power law. (**bottom**) Energy of magnetic neutron resonance measured at a commensurate $(\pi,\pi)$ wavevector in several FeSC systems including 122 [99] and 11 [100, 101] compounds, plotted as a function of $T_c$. Solid line is a linear fit to all data, indicating that the resonance energy scales linearly with $T_c$.

For amplitude-sensitive experiments, the situation is more conflicting among different experiments and it will likely require more work to resolve all the differences. NMR experiments were the first to probe the OP symmetry by considering the temperature dependence of the nuclear relaxation rate $1/T_1$ below $T_c$, with a variety of results dominated by $\sim T^3$ behavior (see [4]), but with exceptions including much lower ($\sim T$) power laws consistent with nodal OP symmetry in BaFe$_2$As$_{2-x}$P$_x$ [106] and higher ($\sim T^5$) power laws more consistent with activated behavior in Ba$_{1-x}$K$_x$Fe$_2$As$_2$ [107]. The intermediate ($\sim T^3$) power laws, while consistent with a 2D line node model, can also arise in a $s\pm$ symmetry model that accounts for disorder scattering, with exponential behavior in $1/T_1$ expected only at low-T [108].

Measurements of thermal conductivity at temperatures approaching absolute zero are one of the best bulk-probe techniques for studying low-lying quasiparticle excitations in a superconductor, with the ability to provide strong constraints on the pairing symmetry [109]. For a fully gapped single-band *s*-wave superconductor, the electronic thermal conductivity follows an activated behavior in both temperature and field dependence due to the full gapping of the Fermi surface, going exponentially to zero with T and H for T<<$T_c$ and H<<$H_{c2}$. For a superconductor with nodes (see **Box 2**), a T-linear electronic component arises due to residual nodal quasiparticle excitations induced by impurity scattering. Also, magnetic field is much more effective in activating thermal carriers, with the Volovik effect imposing a strong enhancement in nodal excitations that increases as $H^{1/2}$. In a multi-band superconductor, particularly with the $s\pm$ symmetry, the exact expectation for thermal quantities is difficult to determine due to the complex interplay of impurity, inter- and intra-band scattering, but has been addressed [110].

Several low-temperature thermal conductivity measurements have now been reported for the FeSCs, with a variety of behaviors observed. **Fig.3** presents a summary of the magnetic field dependence of thermal conductivity measured in several Ba-based 122 compounds in the low-temperature (T=0) limit, plotted as a function of reduced field to display both the T-linear component (H=0 intercept) and the field dependence in the T=0 limit [111-116]. As shown, even in this constrained set of compounds there is a widely varying set of behaviors, spanning the range between the standard *s*-wave and *d*-wave expectations (dashed lines) described above. To date, this is one of the most puzzling results from OP symmetry-sensitive experiments: even within a controlled substitution series, such as in BaFe$_{2-x}$Co$_x$As$_2$, there is a strong evolution of the field enhancement of thermal carriers with doping suggestive of an evolution of the gap structure [112]. Even more perplexing is the fact that, in contrast to Co- and Ni-doped BaFe$_2$As$_2$, a T-linear term is observable in BaFe$_2$(As$_{0.67}$P$_{0.32}$)$_2$ [116], signifying that the development of nodes in the gap structure is sensitive to the type of chemical substitution. Furthermore, the "extremely overdoped" stoichiometric compound KFe$_2$As$_2$ also appears to show low-energy excitations consistent with line nodes [115] as opposed to Ba$_{1-x}$K$_x$Fe$_2$As$_2$ near optimal K concentration [111].

However, this apparently confusing situation may be consistent with the proposed $s\pm$ symmetry due to the development of accidental nodes (or at least deep minima) in the OP under certain conditions. For instance, it has been shown theoretically that disorder can "lift" existing nodes in an $s\pm$ symmetry OP due to the averaging of anisotropy by intraband scattering [117]. It was also suggested that nodes can appear when a third hole pocket at ($\pi,\pi$) is absent [118], or more generally when variations in the electronic structure are imposed by its sensitivity to the Fe-As tetrahedral bond geometry. In the latter case, Kuroki *et al.* claim that the tendency toward magnetism becomes stronger with increasing pnictogen height above the Fe plane, causing multiple spin-fluctuation modes to conspire to result in either a fully-gapped OP or one with nodes [33].

Mishra *et al.* have calculated both the T- and H-dependence of thermal conductivity for several combinations of two-band *s*-wave symmetries, including non-sign-changing and sign-changing cases with anisotropies and varying rates of disorder scattering imposed. They find a plethora of expectations, from fully activated behavior in an isotropic two-gap ($s_{++}$ or $s\pm$) state with no disorder, to a finite non-universal residual electronic term in an anisotropic two-gap ($s\pm$) state that depends non-monotonically on disorder. They note that a strong field dependence at small H rules out a clean $s\pm$ state for 122 materials but is consistent with pair-breaking scattering. However, the pair-breaking would have to be unphysically strong (note that Ba$_{1-x}$K$_x$Fe$_2$As$_2$ and BaFe$_{2-x}$Co$_x$As$_2$ likely have largely different levels of disorder due to ion replacement location). Therefore the conclusion is that the 122 materials likely have a multi-band A$_{1g}$ (*s*-wave) symmetry state with a highly anisotropic gap on one Fermi surface [110]. While this appears to be inconsistent with ARPES experiments, which so far show only isotropic gap structures [35,36,66], the situation may be reconciled in a scenario where the SC OP has a strong 3D momentum dependence. In particular, very recent thermal conductivity experiments on BaFe$_{2-x}$Co$_x$As$_2$ in the zero-temperature limit have uncovered a finite residual T-linear component when thermal currents are directed along the crystallographic *c*-axis [119], which is consistent with the presence of nodes on some part of the Fermi surface. Because the normalized (relative to the normal state) residual conductivity quickly rises for both current directions with field and even becomes isotropic at ~25% of H$_{c2}$, Reid *et al.* suggest a scenario where an effectively 2D Fermi surface harbors a gap with non-zero minima and a more 3D surface (that is warped enough to contribute to *c*-axis transport) harbors a gap with nodes. While the experiment cannot distinguish between a structure with nodes positioned at zero- or finite-valued **k$_z$**, theoretical predictions for a proximity-induced gap would be consistent with *c*-axis nodes [120]. What is most striking, however, is that the residual conductivity is most pronounced *away* from optimal doping, which is either an extrinsic effect due to an impurity- or disorder-induced normal-state contribution (possibly related to the residual heat capacity discussed below), or an intrinsic signature of the nature of the SC gap. As pointed out by Reid *et al.*, "modulation is a sign of weakness", which translates to the provocative idea that deviations from two-dimensionality may weaken the SC state in the iron-pnictides. Nevertheless, future experiments should shed light on these issues. Specifically, magnetic field-angle-resolved thermal conductivity and specific heat studies should delineate the position(s) of nodes and the OP symmetry. Recent specific heat measurements on the related material Fe$_{1+x}$Se$_{0.4}$Te$_{0.6}$ have indeed revealed a four-fold oscillation in the basal plane consistent with either nodes or deep minima in the gap structure [121].

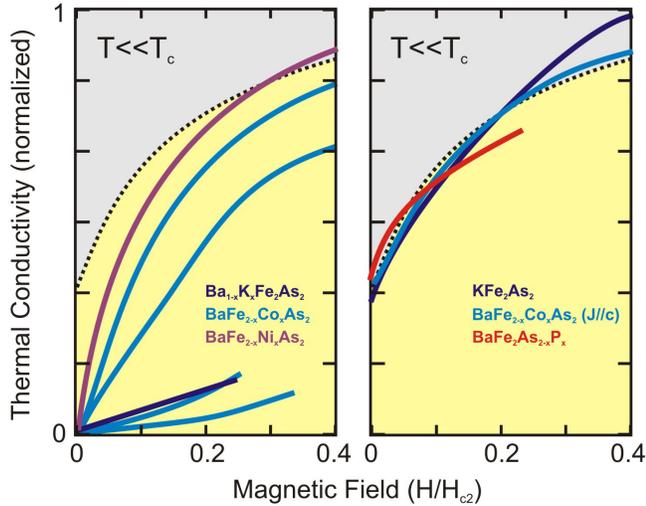

**Figure 3**: **Electronic thermal conductivity of FeAs-based systems in the superconducting state.** Thermal conductivity data in the T=0 limit of several members of the chemically substituted BaFe$_2$As$_2$ family is plotted normalized to normal state values (*i.e.* at H$_{c2}$) as a function of applied field normalized to upper critical field values. Experimental data sets include Ba$_{1-x}$K$_x$Fe$_2$As$_2$ [111], KFe$_2$As$_2$ [115], BaFe$_{2-x}$Co$_x$As$_2$ for basal plane [112] and *c*-axis current orientations [119], BaFe$_{2-x}$Ni$_x$As$_2$ [114] and BaFe$_2$As$_{2-x}$P$_x$ [116]. These are separated for clarity purposes into (**left**) data sets which do not exhibit a residual (i.e. *y*-intercept) contribution attributable to low-energy quasiparticle excitations at zero field, and (**right**) data sets which do show a clear residual contribution at zero field. Note that all systems that exhibit a residual contribution (right) also show exceptionally similar field dependence, closely following the expectation for an unconventional superconductor with line nodes in its gap symmetry (dashed line).

Temperature dependent penetration depth experiments have also given a wide disparity of results, including activated (exponential) behavior in PrFeAsO$_{1-y}$ [122] and SmFeAsO$_{1-x}$F$_y$ [123], and power law behavior in 1111 [124] and 122 [116,125,126] FeSCs and in chalcogenide- [127] and phosphide-based [128] systems. These variations can possibly be explained as due to impurities or other disorder, since the T dependence of the penetration depth is very sensitive to disorder and paramagnetic contributions. This was shown systematically in the 122 system, where the exact temperature dependence was shown to vary as a function of intrinsic disorder in the case of Ba$_{1-x}$K$_x$Fe$_2$As$_2$ [126] and as a function of disorder imposed by irradiation in the case of Co- and Ni-doped BaFe$_2$As$_2$ [129]. Also, the activated behavior in Pr- and Sm-based 1111 materials is put into question in a study by Martin *et al.* on La- and Nd-based 1111 crystals, where the subtraction of a paramagnetic contribution reveals a quadratic temperature dependence of the London penetration depth [124]. This corrected data can be fit with a two-gap *s*-wave model, but issues with the extracted $\Delta$/k$_B$T$_c$ ratios being smaller than the BCS expectations and deviations of the fits near T$_c$ call for a model that considers interband scattering in more detail. Another example of a two-gap fit is shown using a local measurement of penetration depth in Ba(Fe$_{0.95}$Co$_{0.05}$)$_2$As$_2$ using magnetic force microscopy and scanning SQUID spectroscopy, which observed a temperature dependence that varies about ten times more slowly than in bulk measurements [130]. These data were well fit with a fully-gapped model with two different gaps, while the bulk measurement is not obviously consistent with a fully gapped model unless disorder is considered.

Recent studies of the low-temperature electronic specific heat ($\gamma$) have added important insights. In the BaFe$_{2-x}$Co$_x$As$_2$ system, a systematic study of the SC portion of $\gamma$ as a function of x has revealed a lack of any strong doping dependence, signifying no major change in the gap structure through the superconducting range of concentrations; Gofryk *et al.* fit their data using a two-gap model, yielding a factor of two ratio between gap amplitudes that does not vary by more than 10% through several Co concentrations [131]. The width of the SC transition at T$_c$, however, does show a significant increase away from optimal doping. This may be in line with the conclusions from thermal conductivity measurements of weakened pairing strength away from optimal doping [Reid]. A more intriguing find involves measurements of $\gamma$ in the T=0 limit in Ba(Fe$_{0.92}$Co$_{0.08}$)$_2$As$_2$ (T$_c$ = 20K) [131] and BaFe$_2$(As$_{0.7}$P$_{0.3}$)$_2$ (T$_c$ = 30K) [132], both of which report a sizeable residual contribution in the SC state that is ~20% of the estimated normal state electronic contribution. This anomalous residual term is similar in nature to that observed in YBCO [133] and CeCoIn$_5$ [134], both well-established to have nodal OP symmetry. However, the field dependence of $\gamma$ in both FeSC systems does not follow the behavior expected to go along with nodal excitations, such as the well-known H$^{1/2}$

## BOX 2: Electronic band structure and pairing symmetry

The manner in which electrons in a solid behave, in the presence of each other and the surrounding ionic lattice, is well captured by one of the staples of condensed matter physics known as band theory. A metal's band structure can convey a simple yet quantitative description of its electronic, optical and structural properties, and is the basis for understanding many exotic phenomena. In metals, the energy states that participate in determining most properties of a material lie in close proximity to the Fermi energy $E_F$, the level to where available energy states are filled (and therefore unavailable) due to Pauli exclusion.

The band structures of the iron-based superconducting materials have been calculated using first principles density functional theory, finding good general agreement with experimental measurements (see main text). The dominant contribution to the electronic density of states at $E_F$ derives from metallic bonding of the iron $d$-electron orbitals in the iron-pnictogen (or chalcogen) layer. These form several bands that cross $E_F$, both electron- and hole-like, resulting in a multi-band system dominated by iron $d$ character. As shown in **Fig. B2a)** for the case of Co-doped BaFe$_2$As$_2$, the electronic structure is visualized as several distinct sheets of Fermi surfaces within the Brillouin zone, each corresponding to a different band that crosses $E_F$.

Instabilities of this electronic structure to both magnetic ordering and superconducting pairing are widely believed to be at the heart of the exotic properties of the iron-based superconducting materials. For instance, in **Fig. B2a)** one can see that a magnetic ordering vector that spans from the center of the Brillouin zone at $\mathbf{k} = (0,0)$ ($\Gamma$ point) to the corner at $\mathbf{k} = (\pi,\pi)$ (**M** point) will easily nest a circle of points on each of two different Fermi surface sheets (*e.g.*, purple and red sheets), resulting in a spin-density wave order that is driven by properties of the band structure.

Superconductivity is another very well known phenomenon that also results in an "ordered" state that has a strong tie to the band structure. The superconducting order parameter $\Delta$, or "gap function", is a complex function with both amplitude and phase that describes the macroscopic quantum state of Cooper pairs. Its amplitude can in general depend on momentum direction and can change sign via its phase component, but in the simplest case is isotropic (*i.e.*, *s*-wave) and therefore a constant value for all momenta. Details of the pairing potential can instill a less simple case that involves a variation of amplitude as a function of $\mathbf{k}$, or even a variation in phase that results in a change in the sign of $\Delta$ that necessitates the presence of zeros or "nodes" that can take on lower symmetries (*i.e.*, *d*-wave, etc.).

**Fig. B2b)** presents three possible scenarios for the superconducting order parameter symmetry in the iron-based superconductors. With the simplest *s*-wave gap symmetry (*i.e.*, with constant phase) widely ruled out by experimental evidence (see main text), more complicated scenarios are required to explain all observed properties. In particular, circumstantial evidence supports a picture where a change in the sign of $\Delta$ must occur somewhere in the Brillouin zone. With multiple Fermi surfaces, this can be realized by positioning a node either away from the Fermi energy (so-called $s\pm$) or directly at the Fermi energy (*i.e.*, *d*-wave). Moreover, a modulation of the gap amplitude can occur such that, even in the *s*-wave case, so-called "accidental" nodes are present on at least some Fermi surfaces, allowing low-energy excitations to flourish even at temperatures much below the energy of the gap.

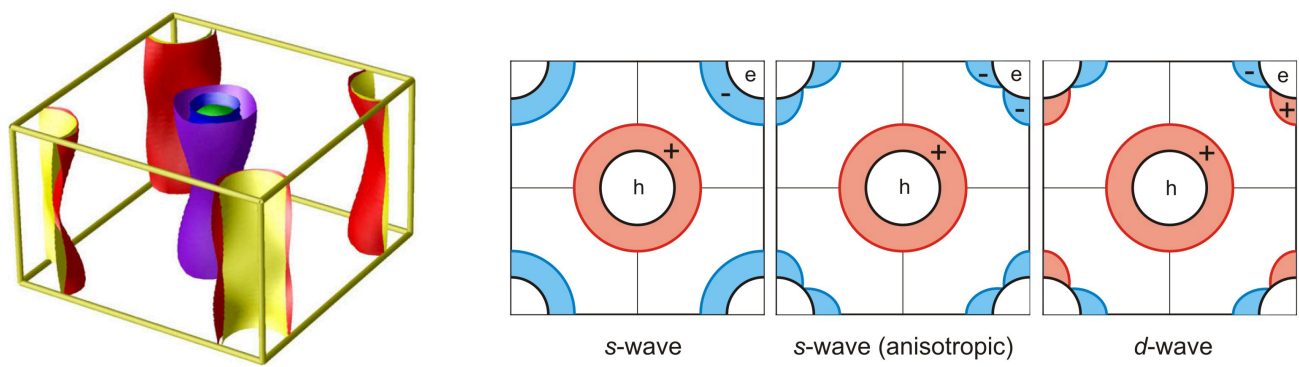

**Figure B2: Fermiology and superconducting order parameter symmetry of 122-type iron-based superconductors. a)** Fermi surfaces of BaFe$_2$As$_2$ with 10% substitution of Co, calculated via density functional theory using experimental atomic positions and drawn using the folded Brillouin zone representation with two Fe per unit cell (from [59]). The hole-like FS pockets (purple and blue) are centered on the $\Gamma$ point [$\mathbf{k} = (0,0)$] and the electron-like surfaces are at the **M** point ([$\mathbf{k} = (\pi,\pi)$). **b)** Schematic picture of the two-dimensional ($\mathbf{k_x}$-$\mathbf{k_y}$) projection of the Brillouin zone of superconducting FeAs-based materials, with multiple bands reduced to single hole (h) and electron (e) pockets. The proposed multi-band pairing gap symmetries, drawn as shaded regions on hole (red) and electron (blue) pockets, are shown for $s\pm$ symmetry with isotropic gaps (left) and anisotropic gaps with accidental nodes on the electron pocket (middle), and for a *d*-wave symmetry (right).

dependence due to the Volovik effect. In Ba(Fe$_{0.92}$Co$_{0.08}$)$_2$As$_2$, a sub-linear field dependence is indeed present but it cannot be fit by a clean *d*-wave gap, and is better described by either a dirty *d*-wave or an anisotropic *s*-wave gap structure [131]. Likewise, in BaFe$_2$(As$_{0.7}$P$_{0.3}$)$_2$ [132] and Ba$_{0.6}$K$_{0.4}$Fe$_2$As$_2$ [135] the observed linear field dependence is more in line with a pairing symmetry that is isotropic on at least one FS sheet that contributes significantly to the electronic DOS [132].

While important, the use of the field dependence of γ as a measure of gap symmetry is notoriously difficult and will require more scrutiny. A more provocative observation has to do with an unusual scaling of the jump in specific heat with $T_c$, as first identified in K-, Co- and Ni-doped BaFe$_2$As$_2$ systems [136]. Because the phonon contribution is difficult to estimate near $T_c$ due to elevated temperatures, it is not straightforward to extract the electronic normal state γ at the transition and thus enable a comparison of the usual normalized ratio $\Delta C(T_c)/\gamma T_c$ to the BCS expectation of 1.43. Instead, Bud'ko *et al.* compared the absolute size of the jump, $\Delta C(T_c)$, and $T_c$ values of several different samples and found a unique power law scaling that appears to go as $\Delta C(T_c) \sim T_c^3$. As shown in **Fig.3**, a compilation of all data available for FeSC systems to date confirms the initial observation to hold in subsequent measurements of K-doped [137] and Co-doped [131,138] BaFe$_2$As$_2$ systems. Measurements in several additional systems, including Rh-, Pd- [139] and Pt-doped [13] BaFe$_2$As$_2$ materials, as well as FeTe$_{0.85}$S$_{0.15}$ [140], LaFePO [141], PrFePO [142], FeSe$_{0.5}$Te$_{0.5}$ [143] and LaFeAsO$_{0.89}$F$_{0.11}$ [144], also fall approximately on the scaling plot. (Note that to date, 1111 FeSCs rarely show a large signature of their SC transitions in specific heat, possibly due to sample quality that may be improved in the future.) Assuming that the enhancement $\Delta C(T_c)$ is purely electronic, this observation is difficult to explain via BCS theory. According to Zaanen, it would require extremely unnatural fine-tuning of the pairing attraction to attain the proper scaling, leading to the conclusion that the underlying ground state is not a usual Fermi liquid but instead is quantum critical [145]. It does appear that, when measured properly, the normal state γ value is enhanced in the higher-$T_c$ materials such as Ba$_{1-x}$K$_x$Fe$_2$As$_2$ [146], lending credence to the incompatibility with a Fermi liquid picture. However, it has also been proposed that strong pair-breaking can also result in such scaling [147]. These opposing views can potentially be elucidated via controlled studies of specific heat and pair-breaking, the latter being discussed below.

First noted in the case of SrFe$_{2-x}$Co$_x$As$_2$ [12], it was quickly recognized that the act of chemical substitution directly into the active pairing layer in the FeSCs presents a stunning contrast to the well-known detrimental effect of Cu-plane disorder in the cuprates. Although there is a consensus that it is better (in the sense of attaining higher $T_c$ values) to minimize disorder within the Fe planes, the mere fact that one can still attain a $T_c$ value near 25 K with strong disorder in the Fe plane (i.e. by Co-substitution) presents a conundrum: how does one disentangle the good (doping/SC optimization) from the bad (pair-breaking disorder)? Pair-breaking studies are extremely useful for identifying a sign-changing OP, where non-magnetic impurities should be strong pair-breakers due to the effect of a scattering event on the momentum-coupled OP phase. However, given the debate about the effects of chemical substitution in FeSCs discussed earlier, it is not clear how any substitution within the active layer can be considered to act only as an impurity scattering center. Luckily, there are other methods of systematically inducing disorder. Karkin *et al.* were the first to study pair-breaking effects in polycrystalline LaFeAsO$_{0.9}$F$_{0.1}$ in a controlled manner using irradiation techniques. They showed a systematic suppression of $T_c$ with disorder induced by fast neutron irradiation consistent with either magnetic defects or non-magnetic pair-breaking involving a nodal gap structure [148]. More recently, alpha particle irradiation was used to study pair-breaking in a controlled manner by irradiating a NdFeAsO$_{1-x}$F$_x$ single crystal sample with $T_c(0)$=46 K [149]. Tarantini *et al.* found that the suppression of $T_c$ by irradiation-induced defects was closely tied to a production of Kondo-like scatting (i.e. $\rho \sim \log(T)$), pointing to defects of a magnetic nature. They claim that the surprisingly weak pair-breaking (in comparison to similar experiments on YBCO) is not compatible with a conventional multi-band BCS picture and is suggestive of a magnetic pairing mechanism. Penetration depth measurements of Co- and Ni-doped BaFe$_2$As$_2$ as a function of irradiation with Pb ions have also shown a "clean" suppression of $T_c$ with increasing defects, as well as a systematic suppression of the low-temperature power law exponent of the penetration depth temperature dependence that is consistent with a nodeless *s*± OP symmetry affected by pair-breaking scattering intermediate between Born and unitary limits [129].

Finally, results from other experiments provide further insight that must be considered. Raman scattering is particularly useful because it is capable of sampling specific regions of the BZ, hence probing the momentum distribution of low-lying gap excitations. A Raman study on BaFe$_{2-x}$Co$_x$As$_2$ samples by Muschler *et al.* provides evidence for a strong gap variation on the electron pockets consistent with the anisotropic *s*± state discussed above [150]. Muon spin-relaxation measurements of the penetration depth of optimally doped Ba(Fe$_{0.926}$Co$_{0.074}$)$_2$As$_2$ fit well to an isotropic two-gap structure with one of the gaps being very small as compared to the BCS expectation [151]. Last but not least, scanning tunneling microscopy experiments on BaFe$_{2-x}$Co$_x$As$_2$ have revealed an unusual gap structure that does not strictly fit either a simple *s*- or *d*-wave shape, but also does not show any four-fold internal vortex structure that would expected in a *d*-wave case [152]. Such direct measures of gap structure will be important for future determination of the true variety of gap structures in the FeSC materials.

**Pairing mechanism, correlations and quantum criticality**

From a general standpoint, the interplay of magnetism and superconductivity strongly suggests that magnetic fluctuations are involved either directly or indirectly in the Cooper pairing in the FeSCs. In the context of magnetism, pairing could arise from fluctuations emanating from a quantum critical point, or an alternative pairing mechanism may simply benefit from the suppression of an antagonistic long range magnetically ordered state. Or a more complicated scenario involving the optimal combination of three players - Coulomb repulsion, spin fluctuations and phonon coupling - may favor high-$T_c$ superconductivity in these materials. First off, phonons alone were quickly ruled out as a standalone contender for the pairing mediator. A calculation of electron-phonon coupling from first principles helped to conclude that LaFeAsOF is intrinsically at most a very poor electron-phonon superconductor [153]. However, a recalculation of this quantity results in a potential increase of the coupling constant by up to 50% when the effects of magnetism are included [154]. The classic isotope effect experiment, which provided the first strong evidence for phonon-mediated Cooper pairing in conventional superconductors, is a well-known method of probing the role of phonons (albeit not always conclusively even for conventional superconductors). Liu *et al.* found an absence of any O isotope effect in SmFeAsO$_{1-x}$F$_x$, but did find a significant Fe isotope effect ("α" exponent of ~0.35) on both the magnetic and superconducting transitions in both SmFeAsO$_{1-x}$F$_x$ and Ba$_{1-x}$K$_x$Fe$_2$As$_2$ [155] suggesting that phonons are at least intermediate players. However, a subsequent study of Ba$_{1-x}$K$_x$Fe$_2$As$_2$ control and isotope-substituted samples grown under identical conditions reports an isotope effect that is much smaller, and even negative [156], motivating the need for more studies.

The strength of the electronic correlations in the iron-based materials is an important issue that has received much theoretical and experimental study. The present experiments show that a modest level of correlations exists, but that the correlations are not as strong as in the cuprates. A simple indication of this is the absence of any clear Mott physics in the FeSCs: the parent compounds are all clearly metallic, and there is no indication of nearby insulating behavior. But conflicting opinions have been generated by experimental results. Optical measurements on LaFePO have indicated that many-body effects are important, identifying this material as a moderately correlated metal [157]. In contrast, x-ray absorption and inelastic scattering measurements on SmFeAsO$_{0.85}$, BaFe$_2$As$_2$ and LaFe$_2$P$_2$ show that their spectra closely resemble that of elemental metallic Fe, thus arguing for weak correlations in these materials [158] (though elemental Fe itself, with a Curie temperature of over 1000 K, is not exactly an uncorrelated system!). In the end, LaFePO may indeed have stronger correlations than 1111 and 122 FeSC materials, which could explain the observations of more unconventional (possibly *d*-wave) SC in this system. Also, in the 122's stronger correlations are certainly nearby. For instance, heavy masses and discrepancies between calculated and observed Fermi surfaces have been observed in quantum oscillations experiments on the "overdoped" compound KFe$_2$As$_2$ [159].

Regardless of the role of strong correlations, spin fluctuations, which possibly provide the collective spin-1 excitation that mediates pairing, are clearly abundant in the FeSCs. Considering the first-order nature of the SDW transitions in these systems, the fluctuations are surprisingly persistent up to rather high temperatures. For instance, the well-known increasing behavior of magnetic susceptibility up to very high temperatures, so far a universal feature of the FeSCs, is considered an indication of AF fluctuations [160]. Also, Mossbauer spectroscopy of BaFe$_{2-x}$Co$_x$As$_2$ has observed a magnetic hyperfine component consistent with short-range AFM correlations at temperatures up to $1.5 \times T_N$ [161], and neutron scattering studies of CaFe$_2$As$_2$ provide evidence for 2D spin fluctuations that persist up to $1.8 \times T_N$ [162].

An NMR study of normal state spin dynamics in BaFe$_{2-x}$Co$_x$As$_2$ has shown that the uniform spin susceptibility is systematically suppressed with increasing Co concentration, with superconductivity optimized near a quantum critical point where SDW order spin fluctuations are maximal [163]. This was suggested to be controlled by the suppression of inter-band transitions (*i.e.* disappearance of a hole pocket) that weakens the nesting condition [54], and it is indeed true that dHvA oscillations observed in the related P-doped BaFe$_2$As$_2$ system show a substantial shrinking of electron pockets and strong mass enhancement upon approach to the magnetic/superconducting boundary at optimal $T_c$ [40]. Because the measured Fermi surfaces in the P-doped series are inconsistent with an estimated (extrapolated) smooth progression between P- and As- end members, Shishido *et al.* concluded that spin fluctuations are the likely explanation for band structure changes, and consequentially superconductivity.

With good ties between superconductivity and spin fluctuations from the suppression of SDW order, it is fair to say that quantum criticality may play an important role. Calculations indeed suggest a competition between electron localization and itinerancy which should yield fertile soil for quantum criticality [164], and the observed specific heat scaling as noted above may provide strong evidence for a truly non-Fermi liquid ground state [145]. However, experimentally there is little evidence for strong deviations from Fermi liquid behavior, especially as compared to other quantum critical systems like the cuprates and heavy-fermion materials. For example, while T-linear resistivity is indeed rampant throughout the optimally doped FeSCs, disentanglement of electron and hole carriers (with some assumptions) has allowed transport measurements of BaFe$_{2-x}$Co$_x$As$_2$ to be interpreted as consistent with Fermi liquid expectations [53]. More important, strong evidence for a Fermi liquid ground state in BaFe$_2$As$_{2-x}$P$_x$ is provided directly by the observation of quantum oscillations throughout its phase diagram [40], even though this system's normal state above $T_c$ shows telltale signatures of a quantum critical point near x=0.66 [165]. Because transport measurements near optimal doping are limited to above $T_c$, it will be an experimental task to develop methods of probing the low-temperature ground state either indirectly in SC systems, or by studying non-SC related materials that can be tuned to criticality, in order to elucidate the true ground state properties near the demise of AFM order. Recent progress has been made to this effect, for instance in Cu-doped BaFe$_2$As$_2$ [16] and P-doped CeFeAsO [82], both of which show complete suppression of magnetic order in the absence of any SC phase.

With difficulties in probing the underlying ground state near the quantum critical point, studies of the coexistence region of SC and AFM are among the most important ongoing efforts. Muon spin resonance experiments have shown contrasting pictures regarding the level of coexistence, with a complete separation of AFM and SC phases evidenced in studies of La- and Ce-based 1111 materials [7] and coexistence of disordered but static magnetism and superconductivity in SmFeAsOF with possible phase separation [8]. In the 122 materials, mesoscopic phase separation into magnetic and paramagnetic (superconducting) regions has been shown in $Ba_{1-x}K_xFe_2As_2$, $Sr_{1-x}Na_xFe_2As_2$, and $CaFe_2As_2$ under pressure [166]. If experiments can show that microscopic coexistence of AFM and SC is indeed intrinsic to FeSC systems, then the competition of these two ground states [96] is not surprising in a scenario where Cooper pairing is mediated by strong fluctuations of AFM order. This is indeed the case in the Ce-based 115 materials, where the same $f$-electron states are known to be responsible for both AFM and SC, with the former giving rise to the latter either competitively as a function of applied pressure [167], or abruptly as a function of chemical substitution [168]. Above all, it is intriguing, and perhaps telling, that commensurate AFM fluctuations appear to be integral to pairing in both the heavy-fermion [169] and FeSC systems.

**Conclusions**

Progress in understanding superconductivity in iron-based materials has advanced tremendously over the last year due to both theoretical and experimental efforts. Considering there is overall good agreement between many different experiments on the general properties of these materials, one can safely conclude that the chemistry is under sufficient control to allow for reasonable comparisons of experimental data without major concerns about sample quality variations. This is one of the main reasons that, after only two years, we have an extensive and reliable set of thermodynamic, transport, surface and spectroscopic data with which to analyze the general and universal properties of this new breed of superconductors. It is also the reason why the observed diversity of properties, especially of the SC state, is so puzzling: the intrinsic nature of these materials appears to include a strong sensitivity to many real and unavoidable perturbations that yields diverse yet reproducible results. Theoretically, this may be due to near degeneracy in energy of different extended $s$-wave and even $d$-wave order parameter symmetry states, implying a sensitivity of gap symmetry to everything from disorder to lattice density. Overall, it is likely that a generic representative order parameter symmetry will involve a sign-changing structure with one fairly isotropic gap and another gap with at least deep minima, and more likely with accidental nodes. In particular, the presence of anisotropic and multi-band scattering, and strong $c$-axis dispersion of at least parts of the Fermi surface structure, will make it difficult to conclude on any particular universal gap structure. But phase-sensitive experiments will help to pin down an intrinsic symmetry, and controlled experiments designed to probe the accidental nature of nodes may provide a better understanding of their dependence on tunable experimental quantities. In the end, these may be moot points, as the relatively high $T_c$ values of the iron-based superconductors do not seem to care much about such details.

Finally, this brief progress article cannot hope to cover all of the ongoing research in such an active and rapidly developing field as the iron-based superconductors. Some important experiments and theoretical ideas have, no doubt, been omitted (for these, the authors express their apologies). For example, the possible role of more exotic physics, such as nematic order [170], has not been discussed. Also, the physics and experiments related to possible applications of these materials have been omitted (for a review, see [171]). The research on the FeSCs has already led to new insights into the novel physics of correlated electronic materials, but there is still much to learn. If the cause of the high-temperature superconductivity in these materials can be understood at the level that conventional electron-phonon superconductors are understood, then it is possible that room temperature superconductivity may prove feasible. Even if this elusive goal is not achieved, the future research on iron-based systems will lead to more knowledge of materials with exotic condensed matter physics properties. As has already been shown by the iron-based superconductors, the cuprates, the organics and the heavy-fermions, superconductivity and magnetism are not as incompatible as was once believed.

**Acknowledgements**

The authors would like to thank Jeff Lynn, Igor Mazin, Doug Scalapino and Louis Taillefer for useful discussions and comments.